\documentclass{aa}
\input psfig.tex
\usepackage{graphicx}
\usepackage{natbib}
\usepackage{array}
\usepackage{graphics}
\usepackage{latexsym}
\usepackage{amssymb}
\usepackage{amsmath}
\usepackage{fancyhdr}
\usepackage{morefloats}
\bibpunct{(}{)}{;}{a}{}{,}

\begin{document}
\title{On the isolated dwarf galaxies: \\ from cuspy to flat dark matter density profiles \\ and metallicity gradients}

\author{S. Pasetto$^{1,2}$, E.K. Grebel$^{1}$, P. Berczik$^{1,3,4}$, R. Spurzem$^{3,5,1}$, W. Dehnen$^6$}

\institute{$^1$ Astronomisches Rechen-Institut, Zentrum f\"ur Astronomie der Universit\"at Heidelberg, Germany \\
$^2$ Max-Planck-Institut f\"ur Astronomie, Heidelberg, Germany\\
$^3$ National Astronomical Observatories of China (NAOC), Chinese Academy of Sciences (CAS), Datun Lu 20A, Chaoyang District, Beijing 100012, China \\
$^4$Main Astronomical Observatory (MAO), National Academy of Sciences of Ukraine (NASU), Akademika Zabolotnoho 27, 03680 Kyiv, Ukraine \\
$^5$ Kavli Institute of Astronomy and Astrophysics, Peking University, Beijing, China \\
$^6$ Department of Physics \& Astronomy, University of Leicester, Leicester, LE1 7RH \\
 \email{{spasetto\char64ari.uni-heidelberg.de}}}
\date{Accepted for publication on A\&A }

\titlerunning{Isolated dwarf galaxies}
\authorrunning{S. Pasetto et al}

\abstract{The chemodynamical evolution of spherical multi-component self-gravitating models for isolated dwarf galaxies is studied. We compare their evolution with and without feedback effects from star formation processes. We find that initially cuspy dark matter profiles flatten with time as a result of star formation, without any special tuning conditions. Thus the seemingly flattened profiles found in many dwarfs do not contradict the cuspy profiles predicted by cosmological models. We also calculate the chemical evolution of stars and gas, to permit comparisons with observational data.
 \keywords{dark matter density profile, dwarf galaxies, chemical evolution, metallicity gradients, Local Group}}

\maketitle

\section{Introduction}\label{Introduction}
Dwarf galaxies are the most common galaxies in the Universe (e.g. \citet{1997AJ....113..185M}) and the Local Group sample is an excellent laboratory where to study these systems and their properties (\citet{1998ARA&A..36..435M, 1999IAUS..192...17G}). Our interest in these structures is related to many important mutually related topics: the growth of structure, the nature of the dark matter and dynamical interactions. In the last decades, new observational data have become available for many dwarf galaxies (in particular in our Local Group) especially regarding the velocity dispersion profiles of stars and the rotation curves of gas, providing clean measures of the dynamical mass at all radii where baryonic matter can be found. This is gradually permitting a deeper investigation on the equilibrium and stability of the dark matter profiles.
Such studies revealed a discrepancy between the prediction of the $\Lambda CDM$
models by numerical simulations, which require power law density profiles with cusps $\rho  \propto r^{ - \gamma } $
 where $\gamma  = 1$
 or $1.5$
 (e.g. \citet{1999ApJ...524L..19M, 1997ApJ...490..493N, 2004MNRAS.349.1039N}),  and the more flattened profiles derived from observations (e.g. \citet{1995ApJ...447L..25B, 2002A&A...385..816D, 2005ApJ...634L.145G, 2005ApJ...634..227D}). 
 Other problems strictly related to the nature of the dark matter and its clustering properties are, e.g., the missing satellites problem (e.g. \citet{1999ApJ...524L..19M, 1999ApJ...516..530K, 1999MNRAS.310.1147M, 1999ApJ...522...82K}),  the triaxiality of the halos for clusters of galaxies inferred from gravitational lensing (e.g. \citet{1998ApJ...498L.107T}) and the angular momentum problem (e.g. \citet{2000ApJ...538..477N}). 
 Despite the many successes of the $\Lambda CDM$
 paradigm, a large number of open questions remain. Among the large body of literature on the nature of dark matter and tests of different formulations of gravity, the work of, e.g., \citet{2001ApJ...556...93B, 2000PhRvL..84.3760S} and \citet{2004PhRvD..70h3509B} provide classical results.
 Our goal in this paper is to extend these studies by examining the interplay between dynamics and star formation in the evolution of dwarf galaxies. Although including star formation makes the modeling more complex, it also allows for comparisons with dwarf galaxies in the Local Group.

Chemodynamical modeling of this type has been productive in the past, particularly in investigations of the effects of winds from high-mass stars and supernova explosions in the ISM (e.g. \citet{2004A&A...422...55P, 2005MNRAS.356..737S, 2005A&A...436..585D}). Given the small dynamical mass inferred for dwarf galaxies, it is surprising that the ISM can remain bound long enough to permit star formation episodes that last for a few Gyr (as found by, e.g., \citet{2005MNRAS.359..985B, 1986AJ.....92...23C, 2002MNRAS.332...91D, 1998ARA&A..36..435M,2003AJ....125.1926G}) giving a typical baryonic matter\footnote{Hereafter we identify baryonic matter with the stellar component; the dark matter will generally be called not-baryonic matter even if its nature could be baryonic as in the case of baryonic-dark-matter objects such as MACHOs or brown dwarfs.} binding energy of $ \sim 10^{53} \text{erg}$ with a total expected number of SN events of $10^3  \div 10^4 $. 

We will focus our attention in this paper on the dwarf spheroidal (dSph) galaxies. These systems are the least massive galaxies known but their velocity dispersions suggest a mass-to-light ratio of up to $100M_ \odot  /L_ \odot  $, making them some of the most dark matter dominated objects in the universe (e.g., \citet{2007ApJ...663..948G}). For these galaxies, the star formation history and chemical composition are very environment dependent, making the realization of a consistent theory of their origin and evolution that combines dynamics, star formation and chemical enrichment more difficult. High-resolution spectroscopy of several dSphs shows a large spread in metallicity (e.g. \citet{2001ApJ...548..592S}) corroborated by photometric studies
 (e.g. \citet{2001AJ....122.3092H,2001AJ....122.2524A,  2002AJ....124.3222B}). Several authors suggested that the ISM of dSph systems could be entirely removed by SNe explosions (\citet{1986ApJ...303...39D, 2004A&A...426...25H, 2002ApJ...571...40M, 2004ApJ...613L..97M, 1997ApJ...478L..21M, 1999MNRAS.309..161M, 2004MNRAS.351.1338L, 1999ApJ...513..142M}).

\citet{2006Natur.442..539M} discuss how star formation processes can successfully act as flattener for the central density cusps, within very short timescales (more details will be given in the following). \citet{2005MNRAS.356..107R} obtain the same result as effect of an external impulsive mass loss event. \citet{2002MNRAS.333..299G} investigate the influence of winds. \citet{1996MNRAS.283L..72N} were one of the first to point out the possibility that feedback mechanisms can turn the central dark-matter cusp into a cored one.

We start our analysis by demonstrating the gravitational stability of the isolated model. Then, once the stellar processes are included, we combine the dynamical and chemical analysis. In Section \ref{ModelApproach} we outline our research approach. In Section \ref{Settingup} we set up the initial model, in Section \ref{Settingup} we present the evolution of the models with and without stellar processes, in Section \ref{chemistry} we add the chemistry analysis and in Section \ref{conclusion}, we discuss our results. 

\section{The modeling approach} \label{ModelApproach}
In this paper we are trying to address the problem of the dark matter profiles in the dwarf galaxies introduced in the previous Section. Despite the complicated processes acting on the true physical systems, different tools have been developed and presented in the literature to follow the chemodynamical evolution of an isolated spherical multi-component self-gravitating model of a dwarf galaxy system. Briefly, we will base our arguments on a chemodynamical code. Our code can follow the time evolution of dynamically interacting smoothed-particles that represent stars, gas and dark matter components of a dwarf galaxy. The code takes stellar evolution and chemical enrichment of the stellar populations and of the gas into account. A full description of the code used can be found in \citet{1999A&A...348..371B, SpurzemBMKLBMKB09}, to which we refer the reader for details on codification and the implementation.
We will use observational constraints in order to tune our models and simulations. We will proceed as follows:
\begin{enumerate}
	\item We first realize multi-component self-gravitating models of a dwarf galaxy composed of a cuspy dark matter density profile, a stellar density distribution and a gas component.
	\item We evolve these models in isolation to check their gravitational stability.
	\item We activate the star formation (SF) processes to see the changes in the dark matter profiles (if any) and vary the physical parameters.
	\item We compare the results to understand the interplay of SF processes and dynamics.
	\item We turn off the stellar evolution processes to check again the dynamical equilibrium of the resulting system. 
\end{enumerate}

\textit{Our goal is to explore how star formation influences the properties of the dark matter density profiles by changing the baryonic density profiles}. 

The same approach will be tested on an initially cored dark matted density profile (see later sections.)

Our focus on \textit{isolated systems} is the \textit{first hypothesis} that makes our modeling not immediately comparable with observations since completely isolated galaxies probably do not exist in nature.

Our first hypothesis leads to the following corollaries: 
\begin{enumerate}
	\item We can assume a `$\gamma$-profile' (see below) that represents quite well both cored and cusped dark matter profiles in the inner part of the dwarf galaxies. The inner slope of a $\gamma$-profile has the advantage of depending on just one single parameter, i.e. $\gamma$. For the gas and stellar density profiles adopted in the starting models we can again assume gamma profiles as good representation of the projected luminosity profile of the dwarf galaxies in the de-Vaucouleurs form (e.g. \citet{1993MNRAS.265..250D,1994AJ....107..634T})
	\item The outer part of the true dwarf galaxy is not expected to drop off as the $\gamma $-models. However, we are considering a synthetic spherical isolated model, so that for Newton's first theorem at any given radius $\hat r$ the dynamics within $\hat r$ are not affected by the mass distribution exterior to $\hat r$. As a consequence we expect \textit{a priori} no differences in the evolution of isolated models regardless of the slope by which our dark matter profile falls off, in particular in the outskirts of the galaxy whenever spherical symmetry is assumed. This permits us to work consistently with the $\gamma $-models.
\end{enumerate}

\textit{We point out that these two corollaries are not true if the first hypothesis is not assumed}. Nonetheless in the literature often different dark matter profiles are used, especially from the family of the King profiles and the Plummer spheres. But, \textit{even if the gravitational force at a given radius is not dependent on the matter distribution outside of that radius, this is not true for the tidal forces}. Therefore, we expect that in real systems, the tidal influence of external galaxies is relevant for the determination of both the shape of the DM density profile \textit{as well as its inner slope }(except in the unlikely case of a dwarf galaxy evolved at rest exactly at the center of a spherical cluster of galaxies). 
The tidal effects are expected to act both by truncating the dark matter profile by tidal compression and by stretching the outer part of the dwarfs when evolved as satellite systems of a primary. 
Thus, changes in the dark matter profiles in the directions of the leading and trailing tails are also expected, or shallower profiles in the outer part of the dwarfs have to be used if we want to introduce our isolated system in the cosmological framework, e.g., \citet{1997ApJ...490..493N} (hereafter NFW). 
However, that is not our objective here (but see the comment in section \ref{evolution}).

\section{Isolated galaxy}\label{Settingup}
  
\subsection{The model}
Our goal is to provide a self-gravitating three-component model of a dwarf galaxy. The evolution proceeds initially without any astrophysical process other than  gravity. This is done in order to check the consistency of the model and to permit us to isolate the different effects in the evolution of a true astrophysical system due to gravitation or stellar evolution. 

In order to choose realistic parameters for our model of a dwarf galaxy, we look at a system that in the literature has been subject of extensive investigation, i.e. the Carina dwarf galaxy. 
Our target is \textit{not }to realize a specific model of the Carina dwarf galaxy, but first to use observationally motivated input physics and then extend our analysis to the dwarf galaxies class of systems (but see \citet{Pasetto2009b}). The wealth of literature data available on this galaxy, starts already in 1983 with studies of the stellar populations of Carina (\citet{1983ApJ...273..530M}). Color magnitude diagrams were used to derive the star formation history \citep{ 1998AJ....115.1840H,1997AJ....114.1458M,1992PASAu..10...83M,1990A&AS...82..207M,1990A&AS...82....1M,2006MmSAI..77..270M,2006cams.book..272M,2004ASPC..310..133M,2004MSAIS...5...65M,2004MmSAI..75..114M,2003AJ....126..218M,2003MmSAI..74..909M,2004MmSAI..75..110R,2003ApJ...589L..85R}). Spectroscopic measurements yielded chemical abundances, e.g., \citet{2006AJ....131..895K,2008AJ....135.1580K}, the M/L ratio, e.g., \citet{2007ApJ...663..948G} and the dark matter distribution, e.g., \citet{2003ApJ...584..541H, 2007ApJ...663..948G}. Proper motion determinations were obtained by \citet{2003AJ....126.2346P} (see also \citet{2004AJ....128..951P}). Dynamical considerations are available for the orbit from timing arguments in \citet{2008ApJ...679..346M} and for Carina's tidal tails in \citet{2006ApJ...649..201M} which make this system well-suited for further orbit investigations (\citet{Pasetto2009b})

We are adopting the following density distribution profile:
\begin{equation}\label{DehnenProf}
\rho \left( r \right) = \frac{{3 - \gamma }}{{4\pi }}\frac{{Mr_s }}{{r^\gamma  (r + r_s )^{4 - \gamma } }}.	
\end{equation}
From the analysis of the CMD of Carina (e.g. \citet{1998AJ....115.1840H}) we infer that the oldest stellar population contributes no more than 15\% of the stellar mass. Moreover from the work of \citet{1993AJ....105..510M} we get a mass to light ratio in the V band of $\left( {\frac{M}{L}} \right)_V  \sim 23\left( {\frac{{M_ \odot  }}{{L_ \odot  }}} \right)_V $. Thus from Table 3 of \citet{2003ApJ...584..541H} we get $L = 0.43 \cdot 10^6 L_ \odot  $, which leads to a stellar mass estimate of $M \approx 23 \cdot 0.43 \cdot 10^6 M_ \odot   = 9.89 \cdot 10^6 M_ \odot  $. If we now assume that the mass lost by tidal interactions could reach an order of 95\%, following e.g., \citet{2003ApJ...584..541H} we get $M_{\text{Car}}  = 2 \cdot 10^8 M_ \odot$. From this initial mass only $1/50$ is assumed to be observable. This baryonic material consists of 15\% stars, 85\% gas. We thus adopt a starting value of $M_{\text{gas}}  = \frac{{M_{\text{dark}} }}{{50}}\frac{{85}}{{100}} = 3.3 \cdot 10^6 M_ \odot  $
  and $M_{\text{star}}  = \frac{{M_{\text{dark}} }}{{50}}\frac{{15}}{{100}} = 5.9 \cdot 10^5 M_ \odot  $ for our reference model. These numbers can be considered as starting values for an orbiting dwarf galaxy \citep{Pasetto2009b}, but here we will work only on an isolated system, thus for us the gas amount retained by the dwarf galaxy after its oldest stellar population is formed will be considered as a \textit{free parameter}. This approach is assumed because the present-day dSphs contain little or no gas, possibly because a large fraction of gas was expelled by the initial SNII component, by tidal interaction with a host galaxy or in general photoevaporation, rampressure, by star formation processes. Depending on its own individual evolution and on its orbital parameters, each dwarf ultimately evolves under different conditions depending on its orbital peri-center passages and a general study cannot be included in the present work. 
  
  The initial temperature for the gas can be inferred assuming a spherical collapse model for the initial dark matter model and a matter power spectrum $P(k)$ at redshift zero compatible with studies from the SDSS, e.g., \citet{2004ApJ...606..702T} or work on the analysis of the Lyman-$\alpha$ forest, e.g., \citet{2002MNRAS.334..107G}. In this case the typical RMS internal velocity of a halo within $M \le 10^9 M_ \odot  $ is less than $20{\rm{km}} \cdot {\rm{s}}^{ - 1} $. This implies that  e.g., hydrogen with a sound speed higher than $20{\rm{km}} \cdot {\rm{s}}^{ - 1} $ has a velocity $c_s $ greater than the escape velocity of the potential well in which it has to collapse. This gas is hence not in the condition to be retained by the dwarf. Therefore we will assume an initial sound speed in hydrogen of roughly around $c_s  \cong 10{\rm{km}}\cdot{\rm{s}}^{ - 1} $ corresponding to a temperature of $T \cong 10000^\circ \text{K}$.

The evolution of the reference model presented here will be used as normalization case for the realization of a Carina-like dwarf galaxy model that is discussed in more detail in Paper II (\cite{Pasetto2009b}). Carina is one of the few dSph galaxies in the Local Group that shows long-lasting star formation and pronounced intermediate-age populations. This mean Carina resembles to much more isolated, low-mass dwarf irregulars and is one of the reasons why we chose it as a reference for the approximate masses of the different galactic components. We emphasize that it is not the intent of this paper to reproduce the star formation history of the actual orbiting Carina dSph. That is the subject of a separate paper. In fact, in our simulations we explore a larger range of stellar, gas and dark matter masses (Table \ref{Table01}) in order to bracket the classical dwarf spheroidal galaxies given in \cite{1998ARA&A..36..435M,1997RvMA...10...29G}.
  
Another parameter on which Eqn. \eqref{DehnenProf} depends, is the scale radius $r_c $. It is interesting to point out that in the analysis of \citet{2003ApJ...584..541H} of the possible extension of the dark matter halo, they deduce that the tidal radius $r_t $
 of the assumed NFW shaped halos can greatly exceed the surface brightness cut-off $R_{co} $, by an order of $r_t  \ge 20R_{co} $ and that the rate of mass loss can vary from 5\% to 95\% depending on the specific orbit chosen. We proceed by tuning the outer profiles for our dwarf model as follows. We can express the radius containing 90\% of the mass, $r_{90} $, as function of the scale radius for the profiles of Eqn. \eqref{DehnenProf} with $\gamma  = 3/2$ by solving the simple Eqn. $4\pi \int_0^{r_{90} } {\rho \left( r \right)r^2 dr}  = {\raise0.7ex\hbox{$9$} \!\mathord{\left/
 {\vphantom {9 {10}}}\right.\kern-\nulldelimiterspace}
\!\lower0.7ex\hbox{${10}$}}M$, which yelds:
 \[
r_c  = \frac{3}{{19}}\left( {27 + 10 \cdot 30^{1/3}  + 3 \cdot 30^{2/3} } \right)r_{90} 
\]
 For our model we adopt $r_t  \cong r_{90} $ \citep{1995MNRAS.277.1354I} and we assume an initial tidal radius 20 times as large as the present one \citep{2003ApJ...584..541H}.
 With these values the stellar and gas profiles have $r_{90}  \cong \left( {0.5 \times 20} \right) \text{ kpc} = 10\text{ kpc}$, leading to a core radius of $r_{c,T_0 }  \cong 0.72\text{ kpc}$ at the starting time $T_0 $ of our simulations\footnote{The definition of $T_0$ is provided in the following section.}.

 Since we are not interested in the external radial dependence of the profile, as previously remarked, we could produce the same outer radial dependence for the dark matter profile by modifying the scale length in order to obtain the same $r_{90} $, radius containing the 90\% of the mass as used for the baryonic profile. However the approximation adopted for massive particles with Plummer spheres imposes a natural limit on the inner resolution radius we are going to analyze. The assumed smoothing length for our model (see below) is roughly  $\varepsilon _{DM}  \cong 0.120\text{pc}$
 depending on the specific model. We will not attempt an analysis within the limiting radius $r_l  = 2\varepsilon _{DM} $ in the following. Nonetheless we need to be able to discriminate between a dark matter profile with $\gamma  = 2$  or $\gamma  = 0$  for radii $r > r_l $. $\gamma$-models with an assumed total mass of $M_{DM}  \cong 2 \cdot 10^9 M_ \odot  $  are almost indiscernible for $r \in \left[ {r_l ,r_{90} } \right]$  for $r_{90}  \cong 20\text{ kpc}$  at our particle resolution.

\section{Evolution}\label{evolution}

We will focus our attention here only on the properties of interest for dwarf galaxy systems, for which the following results hold. \textit{The role of the triaxiality and the extension to the more general class of elliptical galaxies is not investigated in the present paper}. A table with the range of masses considered here is presented in Table \ref{Table01}. Within this range we mostly sampled models with an $\gamma_{T_0} \equiv \left[M_{\text{dark}}/M_{\text{star}}\right]_{T_0} \in \left[16,688\right]$. This range of ratios partially excludes very massive `dwarfs', e.g., Large-Magellanic-Cloud-sized proto-systems (unrealistic for the scale radius adopted here but see \cite{Pasetto2009c}) and we completely excluded system with initial  $M_{\text{star}}+M_{\text{gas}} \geq M_{\text{dark}}$.

The age of the Universe is 13.7 Gyr, or if expressed as lookback time, $T_{ini}=-13.7$ Gyr. We start our simulations much later, namely at a time when the dwarf galaxies we want to analyze have already formed their old populations.
Our final goal is to study the influence of star formation on the dark matter density profiles. In oder to isolate more clearly this effect we need to neglect the Hubble flow, and all the heating/cooling processes that influence the star formation at hight redshift. Hence, by starting at lookback time of $T_0=-9$Gyr we can follow the phase-space description of our models in a simpler way and implement the star formation processes and their feedback on the interstellar medium with higher numerical resolution. 
Nevertheless, this approach forces us to introduce a preexisting stellar population in order to take into account the gas evolution from $T_{\text{ini}}$ to $T_0$. In this initial time the $\Lambda CDM$ paradigm predicts that the dwarf galaxies are the first structures to form as scale invariant perturbations having masses greater than the Jeans mass of isothermal perturbations at the time of recombination (masses of $ \cong 10^{5 \div 7} M_ \odot  $). Hence, in this scenario we need to consider the time to violently relax the dark halo component within which the proto-cloud will congregate and produce the first stars. 
It is clear that these complicated phases, treated in cosmological simulations, make it more difficult to clearly evidence and disentangle the single stellar evolution influence on the whole gravitational potential of a single system because two fundamental conditions we want to use are missing: \textit{equilibrium} and \textit{isolation}. 
After this initial phase and after an equilibrium configuration has been reached, say at $T_0=-9$Gyr, we expect that an old stellar population has already formed.

\begin{table}%
\centerline{\begin{tabular}{lccl}
\hline
\hline
Total Mass $[M_{\odot}] $ & from       & to  & $\gamma$\\
\hline
$M_{\text{star}}$         & $5.1{\bf{ \times }}10^4 $ & $1.0{\bf{ \times }}10^8$ & $\gamma=1.5$ \\ 
$M_{\text{dark}}$               & $7.1{\bf{ \times }}10^7$ & $1.9{\bf{ \times }}10^9$ & $\gamma \in [0.5,2]$ \\
$M_{\text{gas}} $               & $5.5{\bf{ \times }}10^4$ & $6.9{\bf{ \times }}10^6$ & $\gamma=1.5$\\
\hline
\end{tabular}}
\caption{Range of masses within which the simulations are run. The scale lengths refer to $r_{90}$ as discussed in the text. The values listed here are our starting values at the initial time $T=T_0$}
\label{Table01}
\end{table}
  
\subsection{The evolution without the star formation processes.}\label{noSF}
Once the model has been realized we evolve the system in isolation in order to test the equilibrium of the system's density profiles. The critical parameter is of course the smoothing length of the particles, although in the treecodes the gravitational force estimation is largely performed through cell-cell interactions. Different studies in the literature can be found on the best-fit tuning of the smoothing length parameter (e.g. \citet{2001MNRAS.324..273D, 2000MNRAS.314..475A, 2006ApJ...639..617Z} and references therein). We mention here that different codes require to be tuned on different softening lengths to grant the stability of the starting model depending on the implementation of the code. The initial conditions evolved on a special purposed hardware (GRAPE) reach stability with $\varepsilon  \cong 0.12\text{ kpc}$.

\begin{figure*}
\centering
\includegraphics[width=17cm]{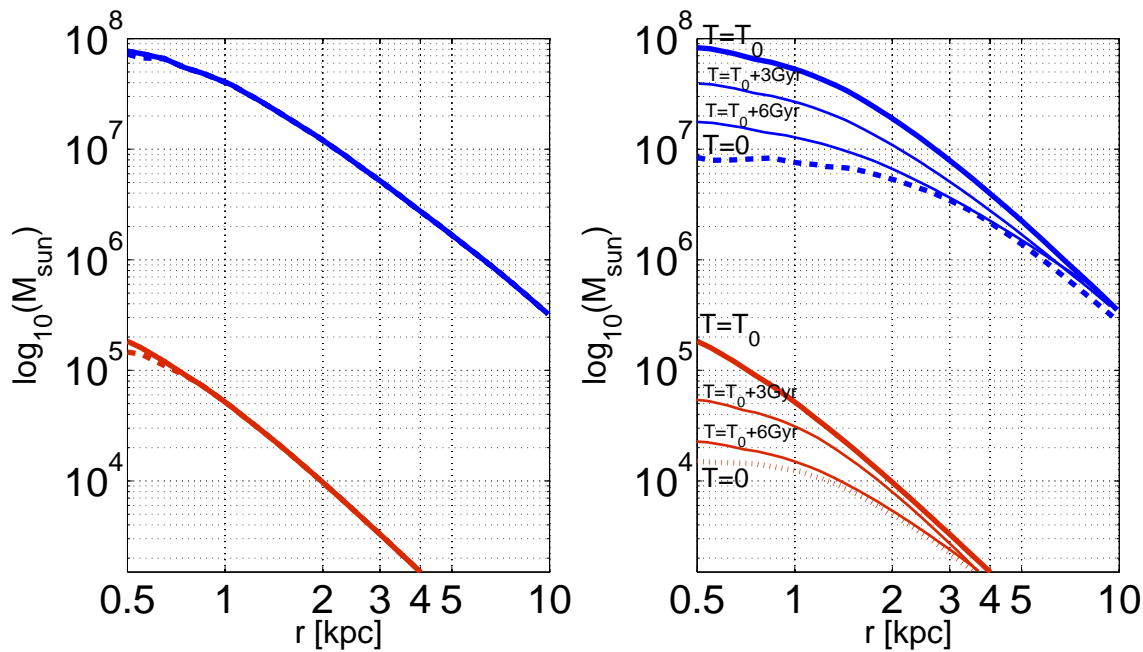}
\caption{(left panel) Density profile temporal evolution for the model evolved in isolation under the action of gravity alone. The blue lines represent the dark matter profile at the beginning (solid) of the simulation and at the end (dashed). The red lines show the stellar profile. No apparent evolution of the dark density profile is seen confirming the equilibrium of the model simulated. The baryonic component remains exactly the same as the starting profile, hence the red dashed line is not visible here since it coincides with the solid red line. (right panel) Density temporal evolution for the same model of the right panel evolved in isolation \textit{and }considering the most relevant star formation processes. The blue solid line represents the dark matter profile at the beginning of the simulation $T=T_0$. The final profile is the blue dashed line. The same line types are used for the stellar profile (red lines). Intermediate thinner lines refer to intermediate evolutionary stages: $T \cong T_0+2.9$ Gyr and $T \cong T_0+6.1$ Gyr The graph scales are chosen to be directly comparable with similar results obtained in \cite{2006Natur.442..539M}.}
\label{DensityTimeAssieme}
\end{figure*}

As can be seen in Figure \ref{DensityTimeAssieme} (left panel), stability is achieved within the resolution used here. The figure represents a simulation with 35000 particles $\left( {n_{\text{gas}} ,n_{\text{star}} ,n_{\text{dark}} } \right) = \left( {1.0,0.5,2.0} \right) \times 10^4 $. The same stability is obtained with slightly different values for the smoothing length $\varepsilon $
 of the dark matter, stars, and gas with a larger number of particles $\left( {n_{\text{gas}} ,n_{\text{star}} ,n_{\text{dark}} } \right) = \left( {2.0,1.0,4.0} \right) \times 10^4 $
 and $\left( {n_{\text{gas}} ,n_{\text{star}} ,n_{\text{dark}} } \right) = \left( {1.0,0.5,2.0} \right) \times 10^5 $
 for a total amount of 350000 particles \textit{in the beginning} of the highest resolution simulation. Even if the higher resolution would permit us to analyze regions closest to the core of the dwarf galaxy, we will see that our chosen distance of $0.5\text{ kpc}$
from the core to which we limit our consideration due to the smoothing length resolution of the dark matter particles, is enough to confirm our results. In Figure \ref{DensityTimeAssieme} (left panel), the overlapping lines represent the evolution of the system at $T = T_0$, i.e. 9 Gyr ago, and after $\Delta T = 9\text{ Gyr}$, i.e. now ($T=T_0+9=-9+9=0$ Gyr), respectively. The density profile of the model hardly changes with time as is desired, both for dark matter profile as for the stellar component, thus confirming that the initial conditions of our model are reasonable.

\subsection{The evolution with the star formation processes.} 
Once the equilibrium of the density profile has been confirmed under the effect of gravity alone, we activate the star formation processes.

The understanding of the star formation processes was a primary task of the Astronomy during all the second half of the last century (e.g., \citet{1969MNRAS.145..405L, 1987IAUS..115..663S}) but, despite the great improvement of knowledge acquired, the micro-physics description still represents a challenge of the modern astrophysics. Great improvement has been achieved within the framework of the Smooth Particles Hydrodynamics star formation algorithms (\citet{1992ApJ...391..502K, 1993MNRAS.265..271N}) that we adopt in our description in its more updated recipe as explained in \citet{1999A&A...348..371B}. 

We emphasized again that we start our simulation \textit{after }the old stellar populations have already formed. For instance, if it took the dwarf galaxy $\approx 3$ or 4 Gyr to accrete gas, to have it collapse and to experience its first measurable episode of star formation, then we assume this equilibrium stage as our initial time of evolution starting from which we analyze the system ($T=T_0$), 3 to 4 Gyrs after the Big Bang. 

The assumed dynamical equilibrium of the model is as previously described in Section \ref{noSF} for the old stellar component, the gas and the dark matter, under the effect of gravity alone. The system can now evolve from an initial time $T_0 $ until the present time. Since we are interested in systems with cuspy dark matter density profiles, we only present their evolution here. Further considerations on flat dark matter profile systems can be found in Section \ref{conclusion}.

Depending on the initial amount of gas and on the total mass of the system, the general behavior of the system is to expel part of the gas due to the effect of SNII that gradually eject the gas to the outer part of the galaxy. Some of the gas is nevertheless conserved in the inner part of the system thanks to the steep potential well. The dwarf galaxy starts as an extremely dark matter dominated system especially in the inner part of the galaxy where the density profile is very cuspy ($M/L>200$ within 5 kpc from the center). In this sense, the steepness of the dark matter profile seems to be a necessity in the initial phase of the dwarf galaxy where just a few percent of the stellar population are formed after the primordial collapse. There are at least three reasons for this:
\begin{enumerate}
	\item If the initial dark matter density profile is not cuspy but flat, the possibility for the dwarf galaxy to survive several pericenter passages as expected for a dwarf like Sagittarius (e.g. \citet{2001MNRAS.323..529H}) becomes more problematic.
	\item If the initial dark matter profile is not cuspy but flat, the efficiency of the SNII to expel the gas is expected to be higher (e.g. \citet{1998A&A...337..338B, 2005MNRAS.356..107R}) thus increasing the amount of primordial gas necessary to reproduce tidally triggered bursts of star formation in galaxies as may be the case in, e.g., Carina (see Paper II). However, this result is strictly dependent on the initial environment in which the gas collapses and we start our simulations only after a time $T_0$ where all these effects have already taken place. 
  \item Cosmological N-body simulations predict cuspy profiles (e.g. \citet{1997ApJ...490..493N}) with an inner slope proportional to $r \propto r^{ - 1} $. The asymptotic behavior for the $\gamma $  models in the inner zones predicts, in the acceptable range studied here, $r \propto r^{ - \gamma } $, including, by the way, the NFW inner slope.
\end{enumerate}

It is difficult to constrain this residual fraction of gas. It seems to be a free parameter that can play a role in the periodic bursts of star formation recently suggested by \citet{2004A&A...422...55P} or may play a role in the dwarf evolving in an orbit where the SFR can be activated by strong tidal interactions.

\begin{figure}
\resizebox{\hsize}{!}{\includegraphics{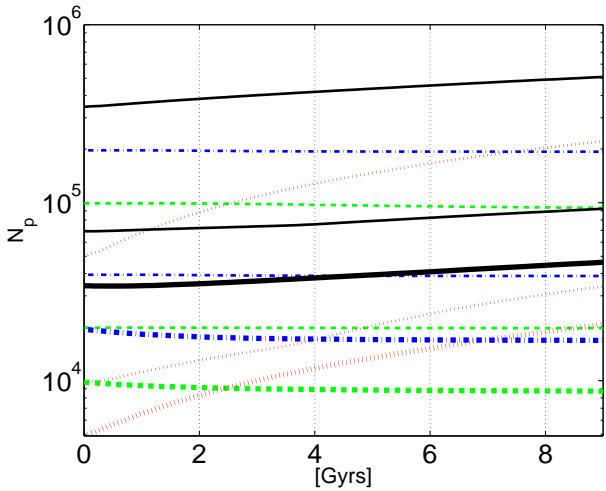}}
\caption{Variation of the number of particles for the reference models and higher resolution test models. Black solid lines refer to the total number of particles, green dashed lines refer to the gas, red dotted lines refer the stars and blue dot-dashed lines refer to the dark matter. The reference model is marked with a thicker line.} 
\label{Npart}
\end{figure}

As we can see from Figure \ref{DensityTimeAssieme} for the reference model of Section \ref{Settingup}, we experience a stellar density profile evolution with the same time scale, $\tau _s $, as the dark matter density profile time scale $\tau _d \cong \tau _s \cong 9 \text{ Gyr} $. The system evolves slowly over timescales of a few Gyr modifying its stellar profile thanks to stellar evolution: new stars are born and die interacting with the small fraction of gas to reach each time new configurations of dynamical equilibrium\footnote{This is obtained whenever the cell opening parameter of the treecode keeps a conservative value, say $\theta  \approx 0.6^\circ $ in order to grant a better energy conservation, or when a more extensive use of the GRAPE hardware is adopted (e.g., \citet{2006sphe.workE..40B,2007sphe.work....5B})}. We point out here how not all the literature recipes for the SF are able to follow efficiently the gravitational changes of the baryonic component. The code really has to `make' the stars such that the number of particles increases regardless of the substantially increasing computational time required. Only these prescriptions, see e.g. Figure \ref{Npart} and \cite{1999A&A...348..371B,2000A&AT...18..829B,2001KFNT...17..213B}, together with the correct gravitational treatment (here provided by a devoted GRAPE hardware) permit one to follow correctly this slow change of the gravitational potential. In Figure \ref{Npart} we plot the range of particles used to test our results and their time evolution. The thick lines refer to the reference models and the thinner ones to the test models as defined in Section \ref{noSF}.  The test model were developed in order to confirm the particle number independence of our results. The slight variation on the number of dark matter particles is due to a small evaporation in the outer part of the dwarf that occurs at distances out of 10 kpc from the center of the system. The number of particles for the stellar and gas components changes depending on the star formation processes (Section \ref{chemistrySF01}) as well as on the evaporation in the outer part of all the components (gas, stars and also dark matter). The similarity of the evolution any trend is remarkably good from the reference model up to the test model with the highest particle number.

Once new stars are added to the stellar component, they reshape the mass distribution (the density profile) and as a consequence the dark matter reacts with the same scale time $\tau _s  \cong \tau _d $ passing thought several states of dynamical equilibrium. 
This can be proven by simply stopping all the star formation processes and evolving the new equilibrium configuration under the influence of gravity alone. The phase-space configuration seems to be stable with respect to the simple dynamical evolution\footnote{This might seem a numerical test on the stability of the equilibrium configurations but it is \textit{not} because it is model dependent. The treecode approach prefers the cell-cell force computation for distant interactions by reducing the role played by the softening length (which is, on the other side, dependent on the growing number of particles used in the systems). The stability of the equilibrium configurations will not be subject of further investigation in this paper.}. 

We observed that the inner flattening of the dark matter halo is accompanied by a simultaneous decrease of the density profile of the baryonic component (Fig \ref{DensityTimeAssieme}). That is simply an indication that the final density profile has to depend \textit{at least} on two parameters (other than the total mass or the central density). As an example, the family of $\gamma $ models or King models permit this kind of metamorphosis: we can obtain this result by simply fixing the radius containing 90\% of the mass $r_{90}  = r_s \left[ {\left( {\frac{{10}}{9}} \right)^{\frac{1}{{3 - \gamma }}}  - 1} \right]^{ - 1} $ and expressing in this way the scale radius as a function of $r_{90} $ and $\gamma $. At fixed $r_{90} $ and total mass $M$ in Eqn. \eqref{DehnenProf}, one can easily obtain central densities that differ by a few orders of magnitude. 

More intuitive is the explanation of the evolution of the stellar density profile because it changes both central density and scale length due to the increasing number of particles and mass, while nonetheless roughly preserving the projected surface brightness. 
\begin{figure}
\resizebox{\hsize}{!}{\includegraphics{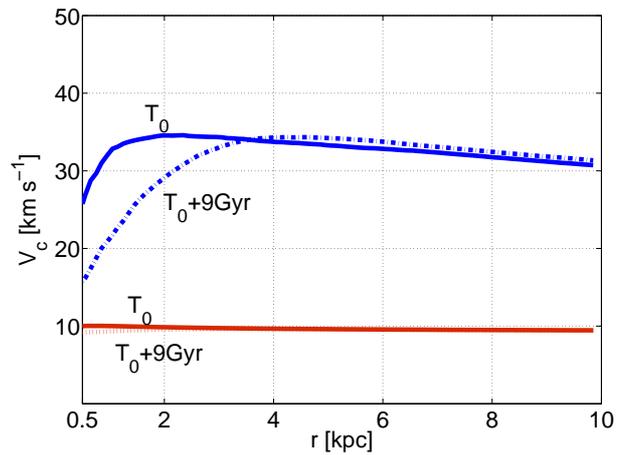}}
\caption{Rotation curve for the dark matter profile (and the stellar component) of the reference model. The solid line represents the starting profile at $T=T_0$ and the dashed line the profile after $T=T_0+9\text{ Gyr}s$. The same colors and line types are used as in the previous figures.}
\label{Vcdark1}
\end{figure}

In Figure \ref{Vcdark1} we plotted the rotation curve $v_c  = \sqrt {r\frac{{d\Phi }}{{dt}}} $ of the dark matter component at the starting time $T = T_0 $ and at the end of the simulation $T = T_0  + 9\text{ Gyr}$ (the intermediate-time snapshots have been omitted for the sake of clarity). The general trend expected as a consequence of the flattening of the profile is exactly the decrease of the central part of the rotation curve, as shown in Fig. \ref{Vcdark1}. We can however observe the small overlap of the rotation curve, which in a two parameter density-potential pair is an indication that the matter distribution at $T = T_0  + 9\text{ Gyr}$ is slightly more concentrated. A fit of the dark matter profile shows that the outer homoeoids for the dark matter profile have slightly contracted. The trend is generally present in our simulations even though the effect here is minimal. Nevertheless, this is a further indication of the fact that a good dark matter density profile has to be at least a 2-parameter model.

\begin{figure}
\resizebox{\hsize}{!}{\includegraphics{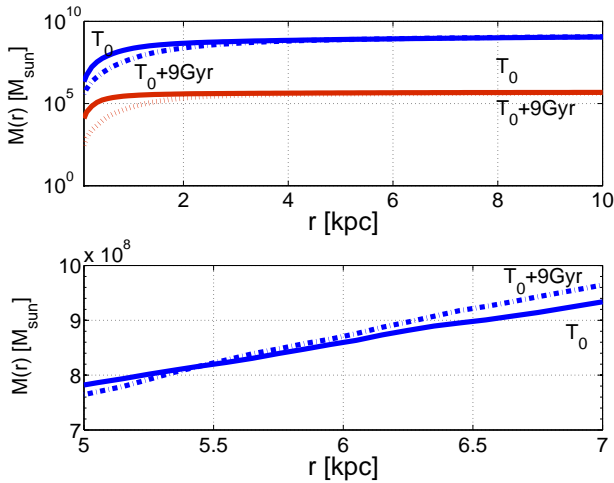}}
\caption{(upper panel) The cumulative mass profile for the dark component (blue) and the stellar component (red). The starting profile is at $T=T_0$ (solid lines) and mostly lower than the profile for time $T=T_0+9\text{ Gyr}$ (dashed lines). (lower panel) A zoom on the dark matter component shows how the lines of the initial and final stage of the evolution cross. See text for details.}
\label{Masscum}
\end{figure}

The cumulative mass density profile $M\left( r \right) = 4\pi \int_0^r {r'^2 \rho \left( {r'} \right)dr'}$ shows this trend as well (Fig. \ref{Masscum}). 
The theoretical expectation for the curves in Fig. \ref{Masscum} indicates that the flatter density profile curve asymptotically converges under the cuspier density curve profile (once we assume the same total mass). 
Considering the asymptotic behavior, the overlapping part of the final cumulative mass profile is an indication of a slightly reduced final scale radius: the same amount of matter is reached in inner homoeoids and starting from a central smaller value as shown in Fig. \ref{Masscum}.
 
Before discussing the kinematics we point out the limits of our simulations. Our finding that a two parameter dark matter density profile is favored, is based on 
\begin{itemize}
	\item the flattening of the density profile at a constant total amount of mass.
	\item the analysis of the rotation curve and cumulative mass profile.
\end{itemize}
But only the first point is stringent evidence for the necessity of the two free parameters model in the dark matter component (in addition to the total mass), while the second point comes with an intrinsic theoretical bias and can \textit{not } be used to claim the necessity to add a further parameter to the NFW profile. The low reactivity of the outer homoeoids to the changes of the inner mass distribution is expected as direct consequence of Newton's second theorem. This theorem describes how the total force acting on the matter distribution out to, e.g., $r_{90}$, is the same as it would be if the shells' matter were concentrated into a point at the center of the dwarf galaxy, despite the different radial distribution it shows in spherical approximation. Thus the finding that the stable dark matter profile has to be at least a two-parameter density profile is just an indication that comes from the baryonic/non-baryonic interaction analysis and does not necessarily rule out the one-parameter density profiles as in \citet{1997ApJ...490..493N}. Moreover as already explained in Section \ref{Settingup}, a model evolved in isolation is expected to show a sharper decline of the external profile as compared to the one resulting from the cosmological collapse of a cube (as we effectively see by comparing a $\gamma$-model with a NFW model).

\begin{figure}
\resizebox{\hsize}{!}{\includegraphics{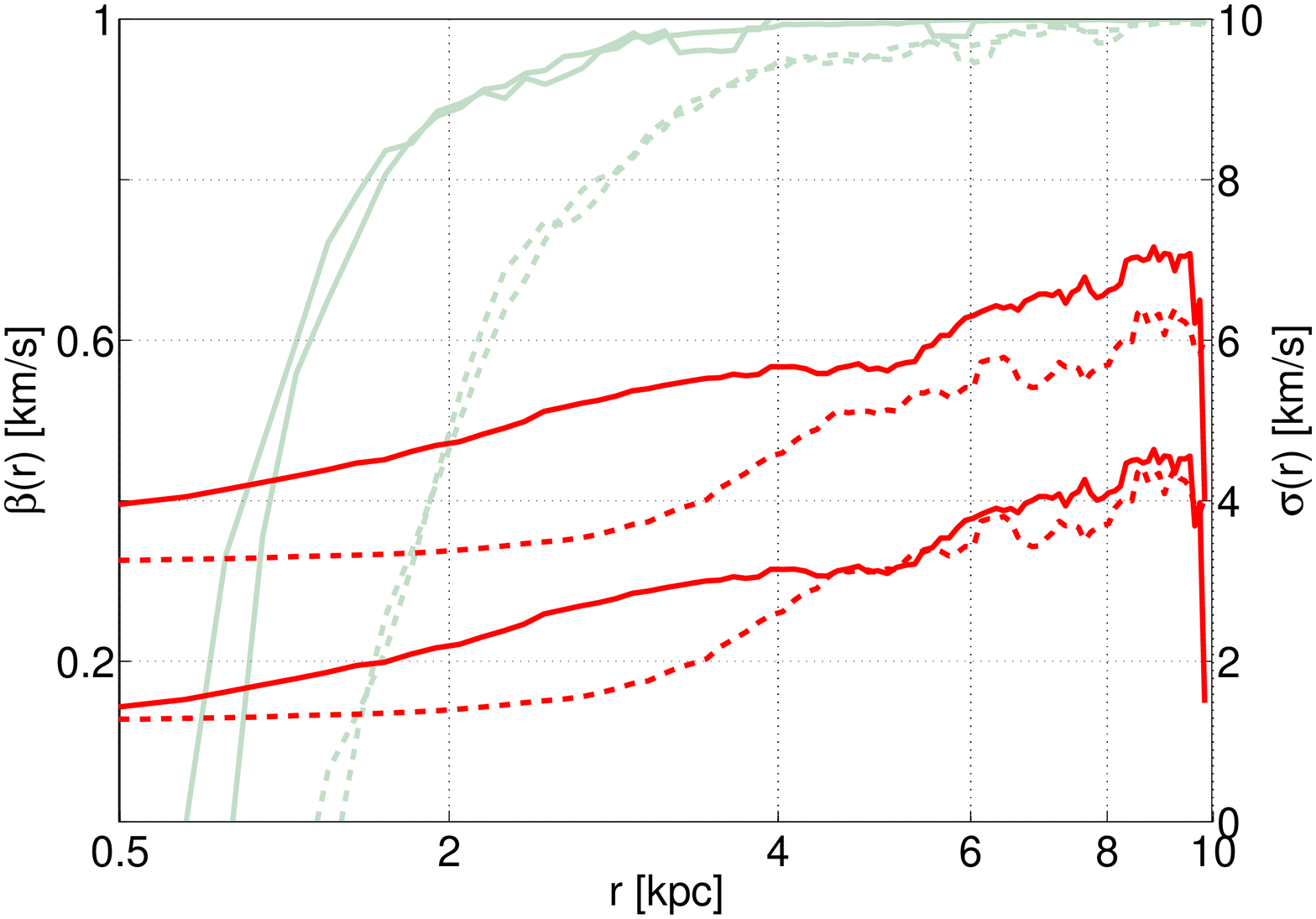}}
\caption{Anisotropy parameter (light green) and radial velocity dispersion (red) for the stellar component. The left scale on the y axis is for the anisotropy parameter, the right scale for the radial velocity dispersion parameter. The thick solid lines show the starting stellar profile, here at $T_0=196$ Myr, and the dotted lines show the resulting profile at $T=T_0+9Gyr$.}
\label{sigmar}
\end{figure} 
 
 To investigate these processes further in detail we present in the following also some chemical considerations that will help us to improve the present theory on dwarf galaxy evolution. The idea to include some chemical considerations, emerges from the analysis of the anisotropy parameter, defined as $\beta  \equiv 1 - \frac{1}{2}\frac{{\sigma _t^2 }}{{\sigma _r^2 }}$, where $\sigma _r$ is the radial velocity dispersion and $\sigma _t^2  \equiv \sigma _\theta ^2  + \sigma _\varphi ^2 $ is the tangential component in a suitable spherical coordinate reference system $\left( {O,r,\theta ,\varphi } \right)$. It is evident from this definition, and from assuming that, as for every reasonable Galactic potential $\sigma _r^2  > \sigma _t^2 $, that the anisotropy parameter $\beta  \in \left[ { - \infty ,1} \right]$, where $\beta  =  - \infty $, corresponds to a system dominated by pure circular orbits $\sigma _r  = 0$, in contrast to a system with $\beta  = 1$ with pure radial orbits. As seen in Fig. \ref{sigmar} the system develops, within the timescale $\tau _d $ of the flattening of the dark matter cusp, a zone dominated by circular orbits, roughly out to a radius $r_{lim}=4$ kpc in radial distance. The kinematics in the outer part of the galaxy are left unchanged in this isolated model\footnote{Any further investigation of the link between the scale length of the starting model $r_c$ and this limiting distance is not the subject of this paper. Nevertheless when evolving this isolated system in an orbit around a host system, we expect that the external tidal force can perturb any relation that may exist between this limiting radius $r_{lim} $ and the scale length $r_c$.} and the radial dispersion velocity profile remains basically unchanged beyond $r>r_{lim}$. Similarly the anisotropy parameter approaches unity at larger radii, getting closer to unity after $r>r_{lim}\cong4$kpc. This indicates how circular orbits become more and more relevant when the dark matter cusp becomes larger, offering a plateau of uniform density and potential, where we expect an easier mixing of the stellar component. This will leave some observable traces in the radial gradient of the chemical composition that we now are going to explain.

\section{Chemo-dynamical evolution}\label{chemistry}
\subsection{Star formation recipes}\label{chemistrySF01}
\begin{figure}
\resizebox{\hsize}{!}{\includegraphics{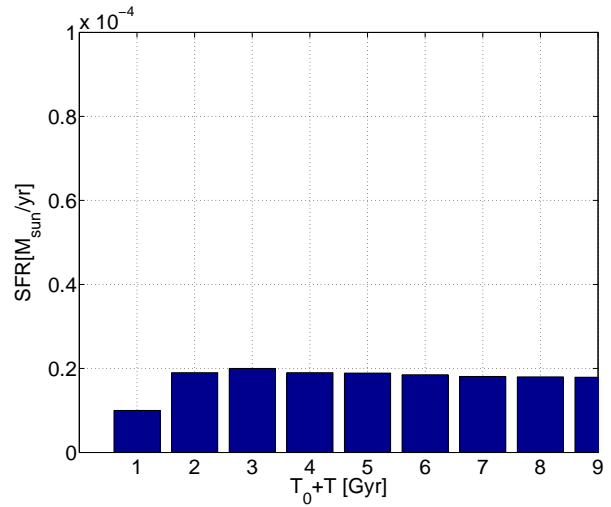}}
\caption{Evolution of the star formation rate with time for the reference model. The total amount of mass (the sum under the area of the histograms), can change by an order of magnitude depending on the free parameter of the amount of gas that the dwarf galaxy can retain, e.g., in our case the initial amount of gas. Nevertheless it is \textit{always } less than a few percent of solar masses per year. The total absence of spurious initial bursts of star formation is also a confirmation of the good level of dynamical equilibrium obtained in Section \ref{noSF})}
\label{SFR}
\end{figure}

The coding of the star formation processes is not the subject of our analysis. A more extended description can be found in the works by \citet{1999A&A...348..371B,2003Ap&SS.284..865B,2003CoKon.103..155B,2002Ap&SS.281..297B,2001KFNT...17..213B,2000KFNTS...3...91B}. Nevertheless the robustness of our evolution has been tested against the different star formation criteria that are  discussed in the literature. 
We found that there are several physical conditions that we \textit{have} to take into account in order to obtain the evolution of the dark matter profiles:
\begin{itemize}
	\item A fluid element has to be instable, e.g., it has to satisfy the Jeans instability criteria $\tau _{sound}  > \tau _{ff}$ between sound speed timescale and free fall timescale.
	\item For the gas particles $\tau _{cool}  \le \tau _{ff} $, the cooling timescale has to be lower than the free fall timescale (this is also often expressed in terms of the over-density criterion $\rho_i  > \rho _{crit}$, i.e., the density of the gas particle $i$, $\rho_i$, has to be higher than a critical density $\rho_{crit}$).
	\item The efficiency of the energy that heats the ISM in a dwarf galaxy can be reduced to a few percent $\epsilon=0.03$, say $\epsilon \eta_{SN I,II}$ where $\eta_{SN}$ is the standard energy injection per SN explosion. This is a result that was derived by the chemical modeling approach already presented in the literature, by, e.g., \citet{1998A&A...337..338B}.
	\item A radius limit of the sphere of influence of a SNII event can be assumed when the sound speed of the front shock is cooled at the sound speed of the ISM. Simulations by \citet{1991ApJ...367...96C} have shown that a reasonable value is of the order of a few hundred parsec.
\end{itemize}
These are sufficient although not necessary conditions. There are other studies that correctly reproduce the observed dynamical effects, e.g., when using dynamical times, $\tau _{dyn}  = \sqrt {3\pi /16G\rho } $ instead of the free fall time in the first criterion (e.g., \citet{1993MNRAS.265..271N}), or when using convergent fluxes with the criterion of the divergence of the velocity $\nabla  \cdot {\bf{v}} < 0$ (e.g., \citet{1998MNRAS.297.1021C}), or even applying higher values of SN efficiency (e.g., \citet{2001MNRAS.322..800R}). 
The first two criteria are universally accepted and do not require any particular fine tuning in our dwarf galaxy simulations to reproduce our results\footnote{ The only slight modification of the Jeans instability criterion is in the case of the presence of magnetic fields. In this situation also the first criterion can be slightly modified in favor of a critical mass depending on the Alfven and the sound speeds. Despite their relevance in the collapse and fragmentation of cold clouds, magnetic considerations have not been investigated in our simulations.} (see Fig. \ref{SFR}).
The last two criteria are \textit{not }tuning parameters. They are results that were already presented in the literature many years ago. They basically act as SN-`shock absorbers' in the dwarf galaxy environment. Thus the result we have presented for the dark matter evolution is \textit{not} dependent on any physical tuned mechanism acting on the baryonic component, nor does it require any `special' mechanism to explain the change of the baryonic density profile and in turn the resulting gravitational effect it produces (the conversion of a cuspy dark matter profile into a flat DM profile).

\subsection{Chemical Evolution}
In our simulations we \textit{assume} that the first episode of star formation and its associated SN explosions led to gas ejection to the outer regions of the dwarf galaxy as described in, e.g., \citet{1999ApJ...513..142M, 2004MNRAS.351.1338L, 2006MNRAS.371..643M}. This is not a direct result of our simulation because, as mentioned, we consider the evolution of a dwarf galaxy after its initial star formation episode which gives rise its old stellar population allowing us to better control the gas present at the initial stage $T=T_0$ of our simulations.

\begin{figure}
\resizebox{\hsize}{!}{\includegraphics{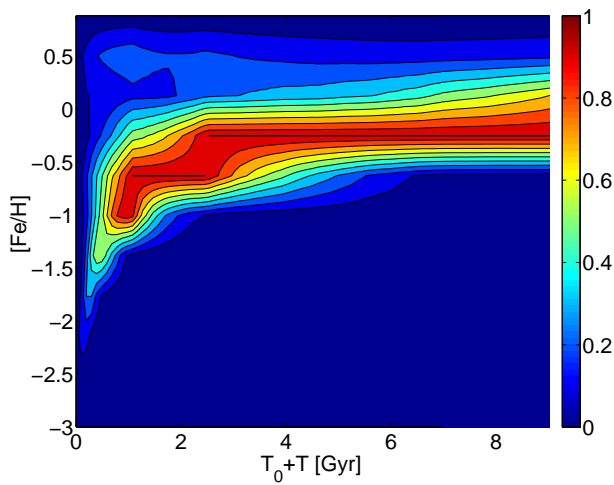}}
\caption{The evolution of the  stellar $[Fe/H]$ over time. The contour plot shows the metallicity spread distribution with time. At the starting time $T_0$ the dwarf starts with a metal poor distribution associated with an old stellar population formed prior to the beginning of our simulation. The color scale is normalized to the highest number of stars (that is not constant in our simulations!) and to the peak of the metallicity per bin. In this way the plot better demostrates the increase of the metallicity with the time \textit{and }its distribution per temporal bin (due to the gradual filling of the distribution tails by SNIa events).}
\label{FesuHT}
\end{figure}

In Fig. \ref{FesuHT} we consider the metallicity evolution with time, without considering its spatial dependence (see Fig. \ref{GradientMet}), by plotting the trend of iron enrichment. Newly formed stars peak above the mean value of iron, $\left\langle {\left[ {Fe/H} \right]} \right\rangle $ which as consequence increases with time and is strongly dependent on the initial chemical composition assumed for the ISM and for the stellar component. We are actually not interested in reproducing any specific observed trend, but we want to understand the expected rate of enrichment. Starting with a rather metal-poor population ($\left\langle {\left[ {Fe/H} \right]} \right\rangle  \cong  - 3$ dex), we are able to see the general rate of enrichment independent from the specific initial conditions that differ from galaxy to galaxy in the true Universe. It is evident that a mean range of  $\left\langle {\left[ {Fe/H} \right]} \right\rangle  \in \left[ { - 1.5, - 0.5} \right]$ can be covered in $9 \text{ Gyr}$ of isolated evolution provided that the system has an initial range of rather low metallicity and then undergoes rapid early enrichment in the initial star formation episodes that need to have taken place during the first, say, 3 Gyr (e.g. \citet{1999ApJ...513..142M, 2004MNRAS.351.1338L}). We note that observationally, such a rapid initial increase does seem to take place, see, e.g., \citet{2007ApJ...657..241K, 2007AJ....133..270K} but we need to remark that a direct comparison with any real dwarf galaxy can not be carried out here. With the trend here evidenced we want to reproduce an 'upper-limit' to the observed chemical enrichment. Thus the results here presented are reference model that can be used to impose an upper limit constraint.

\begin{figure}
\resizebox{\hsize}{!}{\includegraphics{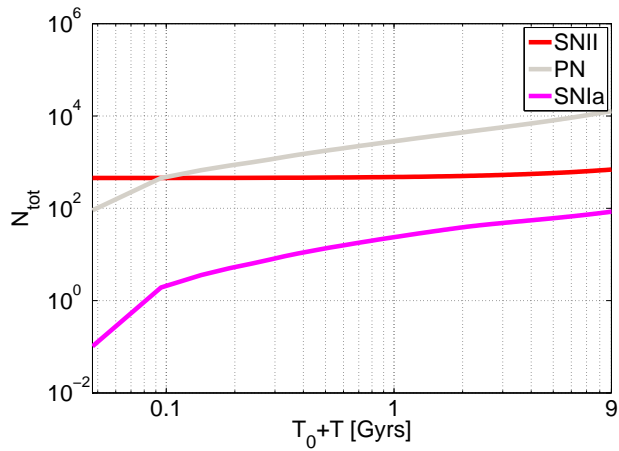}}
\caption{Cumulative number of SNII events in red, SNIa in magenta and planetary nebulae in gray. The number of SNII is almost constant due to the assumption of continuous star formation after the initial collapse which is assumed to to have occurred prior to time $T_0$, by creating the old stellar population of our dwarf galaxy.}
\label{SNs}
\end{figure}

In Fig. \ref{SNs} we plot the number of SN events and PN as a function of time. This helps to explain how the spread in the previous figures originates. In line with the arguments of different authors (e.g. \citet{2006MNRAS.371..643M}) we see how the SNIa, mainly responsible for the iron enrichment of the ISM, operate to fill gradually the tail of the $(\left\langle {\left[ {Fe/H} \right]} \right\rangle-t)$ diagram, even though their number is only a few percent of the SNII. This can indeed be deduced from the Figure \ref{SNs} where the number evolution of SN events is presented as a function of time. Depending on the chemical enrichment we assumed for the oldest stellar population, at the end of the 8 to 9 Gyr of evolution we can obtain a mean present day metallicity of  $\left\langle {\left[ {Fe/H} \right]} \right\rangle  =  - 0.5$ integrated along the radial direction (cfr. \citet{2004MNRAS.351.1338L, 2006MNRAS.371..643M}).

\begin{figure}
\resizebox{\hsize}{!}{\includegraphics{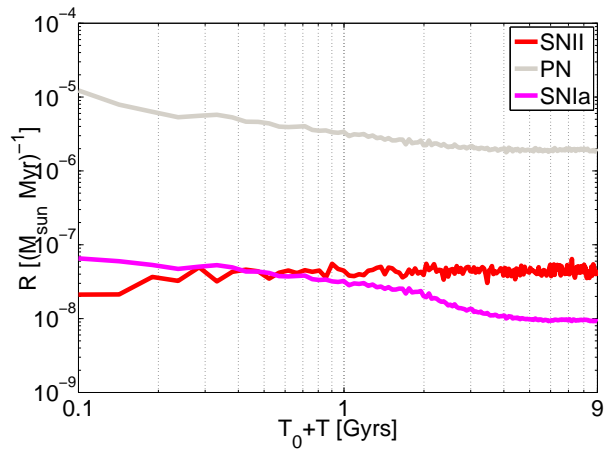}}
\caption{Here the SNIa rate is presented. We find agreement with the literature prediction (e.g. \citet{1983A&A...118..217G}) of a decreasing SNIa rate with time (see also \citet{2001ApJ...558..351M})}
\label{SNsrate}
\end{figure}

The initial dominance of SNII is partially suppressed in this model thanks to the ad-hoc condition avoids the initial burst of star formation (see Fig \ref{SNsrate}) and the SNIa rate is decreasing with time as, e.g., predicted in  \citet{2001ApJ...558..351M}). The reason is that in the collapse of the proto-cloud the star formation is enhanced by the increasing density ($SFR \propto \rho ^{1 \div 2} $) thus favoring the SN events, while in our simulation this phase is neglected and a starting Population II sample is always initially present at $T=T_0$.

\begin{figure}
\resizebox{\hsize}{!}{\includegraphics{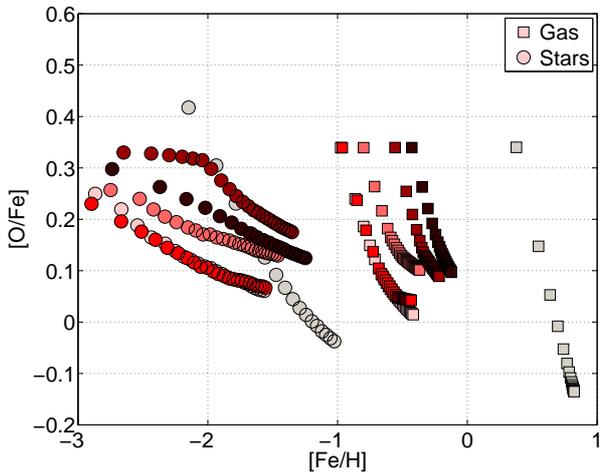}}
\caption{The evolution of the $\alpha$ elements (represented by oxygen) and iron for different initial gas fractions. We show the evolution of the mean $[Fe/H]$ and $[\alpha/Fe]$ in the gaseous and stellar components. The characteristic flat trend of $[\alpha/Fe]$ at low metallicities is not seen here since our simulations start after the first significant episode of star formation. The total mass of the gas is incremented from dark to lighter red: $M_{\text{gas}}=5.81 \cdot 10^5 M_ \odot $, $M_{\text{gas}}=6.55 \cdot 10^5 M_ \odot $, $M_{\text{gas}}=7.31 \cdot 10^5 M_ \odot $, $M_{\text{gas}}=3.23 \cdot 10^6 M_ \odot $ and $M_{\text{gas}}=6.46 \cdot 10^6 M_ \odot $. The supersolar evolution comes from a test model with $M_{\text{gas}}\cong 12.8 \cdot 10^8 M_ \odot $ inside the usual scale length of $r_{s,gas}=0.51$ kpc (see Section \ref{conclusion})}
\label{FesuOmulti}
\end{figure}

Prior to the onset of contributions of SNIa, stellar metallicities ($[Fe/H]$) tend to increase while the $[\alpha/Fe]$ ratio remains rougly constant. This leads to a flat, plateau-like feature at low metallicities when plotting $[Fe/H]$ vs. $[\alpha/Fe]$ (\citet{2004MNRAS.351.1338L, 2006MNRAS.371..643M}). Since our simulations only start after the formation of the old stellar population, our Fig. \ref{FesuOmulti} does not show such a plateau. Instead we see the characteristic decline of $[\alpha/Fe]$ with increasing $[Fe/H]$ due to the combined action of Fe production in SNIa and $\alpha$ element production in SNII.
  
 Moreover, differences in the initial gas fractions affect the metallicities reached during the first episodes of star formation. Lower gas fractions imply less enrichment. Fig. \ref{FesuOmulti} shows results for different initial gas fractions of $15\%$, $30\%$ and $45\%$ as an example. These result in differences in the amount of gas that the galaxy is able to retain at a fixed dark matter amount. Of course, the true behaviour of a dwarf galaxy depends on its intrinsic properties but also on the tidal interaction that may activate SF processes.
 
\subsection{Radial chemical gradients} 
We now turn to the radial dependence of the chemical composition. The spatial distribution of stellar populations of different ages and metallicities is a matter of debate in different studies (see below). Here we want to show what is expected from the simple evolution of our isolated synthetic model. 
Observationally one tends to find gradients such that the younger, and/or more metal-rich populations are more centrally concentrated (e.g. \citet{2001AJ....122.3092H,2006AJ....131..895K}).
The general expectation is that the SNe generate local enrichment at the beginning of the evolution, and that then diffusion tends to erase any radial gradient with time. This affects both iron and oxygen but on different time scales. While oxygen is produced by SNII and so distributed across the entire dwarf galaxy quite fast, iron is expected to be more concentrated in the central regions where the SNIa are more densely located (e.g. \citet{2006MNRAS.371..643M}). As result the gradient is expected to be stronger for iron and longer lasting, while smaller and shorter in duration for oxygen.

\begin{figure}
\resizebox{\hsize}{!}{\includegraphics{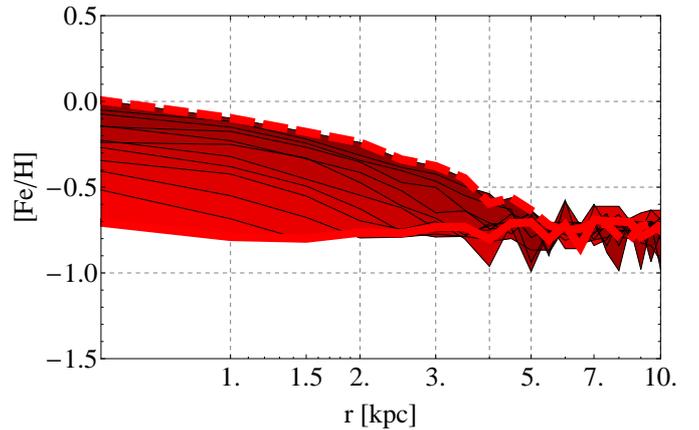}}
\caption{Radial dependence of the mean $[Fe/H]$ value. $[Fe/H]$ is averaged along spherical shells. In red we have plotted the stellar metallicity radial gradient from $T=T_0$ (thick line) to $T=T_0+9$Gyr (the dashed line) with steps $\Delta T= 1$ Gyr . }
\label{GradientMet}
\end{figure}

In this analysis we suggest a possible link between the timescale of the change of the dark matter density profiles and of the formation of chemical gradients which also involve the evolution of the stellar orbit families. If we consider Fig. \ref{GradientMet} 
together with the evolution of the anisotropy parameter, in Fig. \ref{sigmar}, we see how the progressive growth of the number of stars and the resulting changes of the stellar density profile cause a change in the overall gravitational potential, leading to a new equilibrium configuration for the dark matter, which changes its profile from cuspy to flat. For the stellar component we can study the evolution of the orbits, which helps us in interpreting the gradient and the link with the dark matter. 
As can be seen in Fig. \ref{GradientMet}, in the first few Gyr, the gradient is present only in the central zones, where the star formation is more intense due to the higher density. The spatial extent of the gradient encompasses up to 1 kpc after 1 Gyr, up to 2 kpc after 3 Gyr etc. This effect is initially due to the higher fraction of radial orbits (Fig \ref{sigmar}) which mix the inner and outer regions of the system with high efficiency. As time passes the new density profile generated by the newly born stars gives rise to a growing fraction of circular orbits (preferred when a system is slowly evolving, because they are closer to the equilibrium and at lower energy). This can also be inferred from the slow decrease of the radial velocity dispersion which leads to a decrease of the anisotropy parameter. This is happening within the limiting radius $r_{lim}$, within which the chemical gradient and the anisotropy parameter develop for the 9 Gyr of our simulation. Once the fraction of circular orbits is increased (corresponding to the decrease of the anisotropy parameter seen in Fig. \ref{sigmar}), then the mixing efficiency decreases, because there are fewer orbits that extend to the galaxy outskirts. This, together with the gradually decrease of the gas fraction, acts to slow the radial expansion of the zones where the chemical gradient can be evidenced: as seen in Fig. \ref{GradientMet} the lines for T=7 Gyr, T=8 Gyr, T=9 Gyr lie closer and closer to the dotted line ($T=T_0+9$ Gyr) for $r<5$ kpc.
Evidence of such a radial gradient is already present in the literature on N-body simulations, see e.g., \citet{2002MNRAS.335..335C} where the massive spectrum of the elliptical galaxy and the role of the stellar winds are further investigated. 

\begin{figure}
\resizebox{\hsize}{!}{\includegraphics{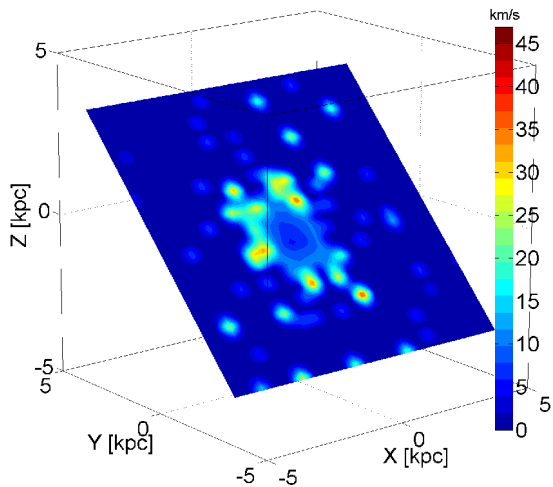}}
\caption{Shells of isovelocity for the sound speed of the gas component. The sketch shows a cut tilted by 45 degrees w.r.t. the vertical axis Z in order to evidence the overlapping shells of isovelocity (see text for details).}
\label{shell}
\end{figure}

In our simulation we did not include galactic winds, thus excluding them as possible explanation of the chemical mixing and consequently a faster disappearance of the $[Fe/H]$ gradient. We point out that recently the same result has been found with a different mechanism, but still avoiding stellar winds by \citet{2008MNRAS.386.2173M}. We agree with the general result presented in their paper and extend it by showing that the reshaping of the dark matter profile seems to work on the same time scale of the metallicity gradient formation. The overlap of the shells of the SNII is also evidenced in Fig. \ref{shell}. In this figure an especially intense episode in the time evolution around $T_0+T=0.730$ Myr is plotted where two SN events are clearly visible (in red) and the remnants of at least other two other ones are still evident (in orange/yellow). Different spots of higher temperature/sound velocity are shown in the velocity surface cut arbitrary tilted by 45 degrees w.r.t. the vertical axis Z. This is a convenient representation of the velocity (or temperature) map that is partially comparable with Fig 3 of \citet{1991ApJ...367...96C} but here in a full 3D description. We clearly see how two green shells of irregular shape (sound speed contour of $25{\rm{km}} \cdot {\rm{s}}^{ - 1} $) from two different SN remnants are overlapping.

On the one hand, we are able to follow the behavior of the orbits and the evolution of the density profiles. On the other hand we can not reach the resolution of the chemical codes used by \citet{2006MNRAS.371..643M} thus preventing us from confirming the general efficiency of the mechanism invoked by these authors. Nevertheless we sustain their complementary picture because we also find the possibility of overlapping polluted regions of the SNII (see, e.g., Fig \ref{shell}). In this figure the isocontours of the sound speed velocity of the gas component are shown. A limiting radius for the sphere of influence of the SNII event feedback on the ISM is assumed when the sound speed of the front shock is cooled at the sound speed of the ISM, here ${\rm{11km}} \cdot {\rm{s}}^{ - 1} $ (see Section \ref{Settingup} and \citet{1991ApJ...367...96C}). This permits it to have overlapping phenomena of the shock fronts in the dwarf galaxies as proposed by \citet{2008MNRAS.386.2173M}.

\section{Conclusions, discussion and limits of the approach}\label{conclusion}

In this paper, we focused our attention on the problem of the dark matter density profiles by investigating the apparent incongruence between dynamical theoretical predictions of a cuspy dark matter profile and the observations that indicate a flatter profile. With our simple description based on the `classical' approach of smoothed particles hydrodynamics (e.g. \citet{1989ApJS...70..419H}) we have obtained the results that we briefly summarize here. 

The classical description of the star formation processes seems to be sufficient to change a cuspy profile into a flatter one. This is shown with the classical smoothed particles approach to the hydrodynamic equations of motion completed with the star formation criteria based on the Jeans theory. The evolution of dark matter profiles could offer a natural explanation for the apparent flat density profiles found in observations (\citet{1995ApJ...447L..25B, 2002A&A...385..816D, 2005ApJ...634L.145G, 2005ApJ...634..227D}) once applied in a realistic context. If we introduce our theory in the context of the cosmological hierarchical merging picture, where the dwarf galaxies are the first stellar population systems that form, the theory developed here predicts that today the star formation processes should have flatted  all the initially cuspy dark matter density profiles (for the range in masses explored). Any dwarf galaxy that contains a stellar population older than, say 8 Gyr, is expected to have a flat dark matter profile, because either the tidal gravitational effect has changed the dark matter profile, or because internal star formation processes acted on the dwarf in isolation as presented here. The combination of both effects is in fact expected to accelerate the profile flattening by tidally activating and enhancing the star formation processes. 
We point out here that the dark matter density profile change happens as soon as the star formation processes act to modify the gravitational equilibrium of the stellar component and, in turn, the gravitational equilibrium of the dark matter. Once the star formation efficiency is reduced, i.e. the stellar over gas fraction exceeds the smaller value predicted in Table \ref{Table01}, then the effect is minor. An initially very gas deficient dwarf galaxy can not experience this profile change. On the other side, galaxies that reach dynamical equilibrium with a high amount of gas still bound to the system (say $M_{{\rm{gas}}}  = 10^7 M_ \odot$ at $T>T_0$) do not have observational counterparts because they produce a dwarf system with super-solar $\left\langle {\left[ {Fe/H} \right]} \right\rangle $ (see, e.g., Fig \ref{FesuOmulti}).
Nevertheless dwarf galaxies with an initial old stellar population of less than $10^4 M_ \odot$ probably belong to a different dynamical regime that is not of interest for us here.

By analyzing the time evolution of the dark matter profiles, we extensively investigated their dynamical properties and evidenced the necessity for at least a two free parameter density model (in addition to the total mass) in order to correctly represent the shape resulting from the baryonic/non baryonic interaction. This is a result partially dependent on the isolation of the model and does necessary rule out NFW profiles, which are obtained only in a baryon-free simulations.

In our simulations, we used an extremely strongly peaked cuspy DM profile in order to study the effect expected in a case of a very deep potential well for the dark matter profile. As consequence these phenomena act even on smaller time scales for a central shallower dark matter density peak, or for small $M/L$ ratios. The same effect, in a different cosmological context, has been described by \citet{2006Natur.442..539M}, although they derived a very short timescale for the effect ($\tau _d  < 800$ Myr). The most plausible explanation for their extremely short timescale $\tau _d$ is an `overheating' by SN of the interstellar medium. The role of the SN feedback (especially the SNII sphere of influence and the $\eta_{SMII}$) are more carefully taken in consideration in our present study as compared to earlier work by other authors. This leads us in a natural way to a longer $\tau _d$, which presents attractive advantages. In the context of dwarf galaxies tidally perturbed by the gravitational field of a primary host (i.e., a major galaxy, or a cluster of galaxies) the existence of deeply embedded systems (e.g. Sagittarius in our MW) could become more problematic whenever the orbital pericenter lies close to the center of the hosting system if $\tau _d$ is as short as predicted by \citet{2006Natur.442..539M}. Our models do not suffer this problem because the cuspy profile in our case is lasting several Gyr and is gradually flattened. As a result we are easily able to save the bulk of the orbiting dwarf galaxy around a primary system \citep{Pasetto2009b}. A second difference is that we treat completely isolated models while the \citet{2006Natur.442..539M} have to work with interacting objects. This offers us the advantage to eliminate spurious perturbations and to have a clean determination of the role of star formation alone. The peculiar orbital path of every different dwarf galaxy when evolved in interaction has a determining role on its SFH and on the density profile evolution. Our modeling of isolated system offers easier analysis and theoretical interpretation of the results.

We tried to link the dark matter profile change with an observable property. A suggested indication in the paper is the chemical radial gradient of metallicity. We performed an experiment that shows how a $\left[ {Fe/H} \right]$ radial gradient may originate, in absence of galactic winds, as simply a result of the baryonic reshaping of the density profiles on a timescale compatible with the dark matter profile reshaping. Thus, in this model, the presence of a radial gradient is always expected  in the evolved dwarf as consequence of a smaller anisotropy parameter and a flat dark matter profile shape.
Finally, we want to point out a few general caveats:
\begin{itemize}
	\item The fraction of gas has generally to be assumed as a free parameter (within the range indicated in Table \ref{Table01}) both in isolated and orbiting dwarf galaxies. We expect that different contents of gas left in the primordial dwarf galaxy can cause different star formation histories especially when the system evolves in orbit around a more massive host (e.g. \citet{2001Ap&SS.276..375M,2003A&A...405..931P}), leading to differences in the baryonic and dark matter phase space re-filling. We start with an already present stellar population formed in isolation. We will show in \citet{Pasetto2009b} that the gas consumption is mostly led by the orbit evolution. 
	\item Another type of observed SFH is the one presented by local systems as Dra and UMi. These systems show long-lasting early star formation episodes that led to considerable chemical enrichment. Obviously a substantial amount of the gas must thus have been retained in this early period (of, say, 13-10 Gyr). When this type of system lies within the well of an hosting system, the major role in the density profiles evolution is surely played by the external environment. Even in absence of gas, the tidal stripping is able to smooth system irregularities and to convert them in dSph (e.g., \citet{2001ApJ...559..754M}). This effect is shown to be efficient also in the MW dwarf companions (e.g., \citet{2003A&A...405..931P}) by morphologically evolving their density profiles. By extending these arguments to the more crowded cluster of galaxies (as Coma or Virgo), we enter the galaxy harassment regime \citep{1996Natur.379..613M} that is a pure gravitational effect on which the literature has still put too few constraints from the star formation and chemical point of view. 
	\item While the trend that leads to the chemical radial gradient is universally evident across the range of masses explored (and also generally present in the more extended sample of the elliptical galaxies, see. e.g. \citet{2002MNRAS.335..335C}) its gradient slope is not. $\frac{{d\left[ {Fe/H} \right]}}{{dr}}$ shows a strong dependence on the initial fraction of gas and stars that is not studied here.
	\item The possible role of the triaxiality of the system is not investigated here. This is expected to play a minor role for the range of masses here investigated. Nevertheless the link between the $r_{lim}$, as defined in this paper, and the scale length of the model $r_c$ for the models here adopted has still to be investigated even for the more general class of elliptical galaxies and is the subject of a forthcoming study (\citet{Pasetto2009c}).
\end{itemize}

\begin{acknowledgements}
We thank C. Chiappini, F. Matteucci and C. Chiosi for useful discussions. We thank the referee G. Lanfranchi for useful comments. Simulations were performed on the GRACE supercomputer (grants I/80 041-043 of the Volkswagen Foundation and 823.219-439/30 and /36 of the Ministry of Science, Research and the Arts of Baden-W\"urttemberg). PB acknowledges the special support by the NAS Ukraine under the 
Main Astronomical Observatory GRAPE/GRID computing cluster project. PB's studies are also
partially supported by the program Cosmomicrophysics of NAS Ukraine.
 RS is supported by the Chinese Academy of Sciences Visiting Professorship for Senior International Scientists, Grant Number 2009S1-5 (The Silk Road Project).
\end{acknowledgements}

\bibliographystyle{apj}
\bibliography{BiblioArt}

\end{document}